\documentclass[a4paper,12pt]{article}
\usepackage{latexsym}
\usepackage{epsfig}
\begin{document}
\title{ Cosmological solutions of time varying speed of light theories}
\author{Mahmood Roshan, Maryam Nouri and Fatimah Shojai\\
Department of Physics, University of Tehran,
Tehran, Iran.}
\maketitle

\begin{abstract}
We consider scalar-tensor theory for describing varying speed of
light in a spatially flat FRW space-time. We find some exact solutions in
the metric and Palatini formalisms. Also we examine the dynamics of
this theory by dynamical system method assuming a $\Lambda$CDM
background and we find some exact solutions by considering the
character of critical points of the theory in both formalisms. We
show that for any attractor the form of non-minimal coupling
coefficient is quadratic in terms of the scalar field $\Psi$. Also
we show that only attractors of the de Sitter era satisfy the
horizon criteria.
\end{abstract}

 \section{Introduction}

There are some ideas that suggest the constants of nature, such as
gravitational constant and the speed of light, should be
time--space dependent. \cite{9,8,bek}. Although the number and the
role of fundamental constants are still debated \cite{7}, there
are different mechanisms leading to varying constants. Brans-Dicke
Scalar-tensor theory \cite{8,1} is the first formulation of a
dynamical gravitational constant as while as a theoretical
explanation of Mach's principle is inherent in it. Another
important physical constant which has attracted considerable
attention recently \cite{9} is the speed of light. Theories with
varying speed of light (VSL) have been firstly proposed by Moffat,
Albrecht, Magueijo and Barrow \cite{10,11} as an alternative to
the inflation mechanism solving some problems of Big-Bang
cosmological models \cite{10,12}. In their formulation the Lorentz
invariance is broken and there is a preferred frame, called
cosmological frame, in which the speed of light is only a time
dependent field. In this frame there exists a pre--set function
\cite{11,bar} representing the speed of light and enters in FRW
equations as an input.

It is a well-known fact  that it is possible to
have a varying speed of light theory and \textit{preserving} the
general covariance  and local Lorentz invariance \cite{13}. The
price that have to be paid for this, is to introduce a time--like
coordinate $x^0$ which is not necessarily equal to $ct$. In terms
of $x^0$ and $\vec{x}$, one has local Lorentz invariance and
general covariance. The physical time $t$, can only be defined
when $dx^0/c$ is integrable.

The most general scalar-tensor action
of gravity which allows for a dynamical speed of light is
illustrated in \cite{14}. This action is previously
analyzed by many authors using the metric approach. Demianski et al
\cite{16} present a class of cosmological models derived from
Noether symmetry requirement. These models describe accelerating
evolution of a
FRW universe filled  with dust matter and exhibit
power-law dependence of coupling and potential to the scalar field. There is also some
\textit{tracking} solutions of this model \cite{17}, in which the
time evolution of the scalar field tracks the expansion
rate of the universe.

In our earlier paper \cite{15} we have found the exact classical
cosmological solutions assuming an exponential coupling between
a scalar field, representing dynamical speed of light, and the geometry, with or without cosmological constant.
Here we shall continue our previous work \cite{15} on
investigating the exact cosmological solutions with varying speed
of light.  In the following sections we shall use it and find it's
exact cosmological solutions for the spatially flat universe. We get the
results firstly for a metric theory. Then we discuss the exact
solutions using the Palatini approach in which the connection and
metric are independent degrees of freedom. In the last section we
examine the dynamics of this theory by dynamical system method
assuming a $\Lambda$CDM background. By considering the character
of critical points of the theory we find some exact cosmological
solutions in both formalisms.

\section{The Model}

The Jordan-VSL action which we use here is the one presented in
\cite{14}:
\begin{equation}
S={1\over 16\pi G} \int d^4x \sqrt{-g} \Bigl(F(\Psi)R - 2U(\Psi) -
Z(\Psi)~g^{\mu\nu}
\partial_{\mu}\Psi
\partial_{\nu}\Psi
 \Bigr)
+ S_m[\phi_i,g_{\mu\nu}] \label{1}
\end{equation}
in which $F(\Psi)=(c/c_0)^4$ and $U(\Psi)$ are arbitrary regular
functions of the scalar field $\Psi$, representing the coupling of
the scalar field $\Psi$ with geometry and it's potential energy
density respectively. $c_0$ is a constant velocity and hereafter we sall put
$8\pi G=c_0^4=1$.
The first part of the above action functional is the gravitational
part, including Ricci scalar $R$ and a dynamical term for the
velocity of light with arbitrary coupling function $Z(\Psi)$. The
latter is the action for matter fields, $\phi_i$,  and doesn't
involve the scalar field $\Psi$, so the matter is minimally
coupled to gravity. As emphasized in the introduction, here we
have assumed that there is a time--like coordinate $x^0$, which is
not equal to $ct$ and thus $dx^0/c$ is not necessarily integrable.
The dynamics of $\Psi$ depends on the functions $F$ and $Z$. But
note that $Z$ can always be set equal to unity by a redefinition
of the field $\Psi$. Therefore only one arbitrary function
remains.

\subsection{Metric approach} In the metric approach the metric and
the scalar field $\Psi$ are dynamical variables. The variation of
the action (\ref{1}) with respect to them gives:
\[
F(\Psi) (R_{\mu\nu}-{1\over2}g_{\mu\nu}R) = T_{\mu\nu}
+ \partial_\mu\Psi\partial_\nu\Psi - {1\over 2}g_{\mu\nu}
(\partial_\alpha\Psi)^2
\]
\begin{equation}
+\nabla_\mu\partial_\nu F(\Psi) - g_{\mu\nu}\nabla_\mu\nabla^\mu
F(\Psi) - g_{\mu\nu} U(\Psi)
\label{2}
\end{equation}
\begin{equation}
\nabla_\alpha\nabla^\alpha \Psi = - \frac{1}{2}{dF\over d\Psi}R
+{dU\over d\Psi}\ \label{si}
\end{equation}
Also the weak equivalence principle holds because the matter fields are minimally coupled to the metric. This implies:
\begin{equation}
\nabla_\mu T^\mu_\nu = 0\ \label{4}
\end{equation}
where the energy-momentum tensor of matter is defined as:
\begin{equation}
T^{\mu\nu}= \frac{2}{\sqrt{-g}}\frac{\delta S_m}{\delta
g_{\mu\nu}}
\end{equation}
In a cosmological context, applying the field equations
(\ref{2})-(\ref{4}) to FRW universe in which the metric has the
following form:
\begin{equation}
ds^2=-{dx^0}^2+a(t)^2(\frac{dr^2}{1-kr^2}+r^2d\Omega ^2)
\end{equation}
leads to the following cosmological equations:
\begin{equation}
3F(H^2+\frac{k}{a^2}) = \rho +{1\over 2} \dot{\Psi}^{2} - 3 H
\dot{F} + U \label{H2}
\end{equation}
\begin{equation}
-2F(\dot{H}-\frac{k}{a^2}) = (\rho+p) + \dot{\Psi}^{2} + \ddot{F}
- H\dot{F} \label{dotH}
\end{equation}
\begin{equation}
(\ddot{\Psi}+3H\dot{\Psi}) = 3{dF\over d\Psi}\left(\dot{H} + 2
H^2+\frac{k}{a^2} \right) -{dU\over d\Psi} , \label{ddotPhi}
\end{equation}
\begin{equation}
\dot{\rho} + 3H~(\rho+p) = 0\ \label{dmatter}
\end{equation}

These are c-variable FRW equations, the wave equation of $\Psi$
field and the conservation law respectively.
$H(x^0)=\frac{1}{a}\frac{da}{dx^0}$ is the Hubble parameter,
$\rho$  and $p$ are the energy and pressure densities of a perfect
fluid considered as matter field and dot denotes derivative with
respect to the time--like coordinate $x^0$. These equations  form
a coupled set of nonlinear differential equations for $H(x^0)$ and
$\Psi(x^0)$. The time--like coordinate $x^0$ is related to cosmic
time by the relation:
\begin{equation}
dt=\frac{dx^0}{c}=\frac{dx^0}{F^{1/4}}
\end{equation}

In cosmological application $dx^0/c$ is integrable and gives the
physical time. Therefore the physical Hubble
parameter $H_{p}(t)=\frac{1}{a}\frac{da}{dt}$ can be evaluated as $H_{p}=H(x^{0}) \frac{dx^{0}}{dt}$.

Substituting $\frac{1}{H(x^{0})}\frac{d}{dx^{0}}$ by $\frac{1}{H_{p}(t)}\frac{d}{dt}$ in (\ref{dmatter}) gives:
\begin{equation}
\frac{d\rho}{dt} + 3H_p(\rho+p) = 0\ \label{dmatt}
\end{equation}
This shows that in this model the conservation equation (\ref{dmatter}) is
valid even in terms of cosmic time. However it is possible to have the non-conservation of energy-momentum
 if we change our model. The way that
the conservation relation changes, highly depends on the model.
 For example in the preferred frame approach \cite{10}, violation of the energy-momentum conservation
 occurs because of appearing a source term proportional to the gradient of c in the conservation equation.

The conditions which one should impose on VSL models are usually inspired
by the cosmological puzzles. In order to solve the horizon problem
of the standard cosmology, one should set  \cite{9}
$\ddot{a}/\dot{a}-\dot{c}/c>0$ for the early universe and also one
has $\dot{a}>0$ . These lead to some constraints on the range of
possible values of integration constants, appeared in the
solutions.

The cosmological solutions of the greatest interest are those for
which the time evolution of the Hubble parameter is proportional
to the inverse of the cosmic time (corresponding to power-law
expansion) or a constant (corresponding to de Sitter  expansion).
Considering a spatially flat FRW universe, we shall distinguish
two cases. A c-dominated universe by which we mean $S_{m}=U=0$.
A (c-$\Lambda$)-dominated universe means that $S_{m}=0$ but
$\Lambda$ is not zero and is of gravitational type.

\subsubsection{$c$-dominated universe:}Putting $ S_m=U=0$ in the
equations of motion, we get two independent equations:
\begin{equation}
3FH^2 = {1\over 2}\ \dot{\Psi}^{2} - 3 H \dot{F} \label{Hc2}
\end{equation}
\begin{equation}
-2F\dot{H} = \dot{\Psi}^{2} +\ddot{F} - H\dot{F}\ \label{dotHc}
\end{equation}
Assuming a power-law dependence for the coupling coefficient $F$,
the above equations have the following solutions:
\begin{equation}
H\sim \frac{1}{x^0} \label{H}
\end{equation}
\begin{equation}
\Psi \sim {x^0}^{\alpha} \label{psi}
\end{equation}
And also the coupling function $F(\Psi) \sim \Psi^2$ which is a
particular case emerged by requiring the existence of Noether
symmetry \cite{16}. The cosmic time is defined as (assuming
$\alpha\neq{2}$):
\begin{equation}
t \sim (x^0)^{1-\alpha/2}
\end{equation}
Thus the physical Hubble parameter and the speed of light can be
written as:
\begin{equation}
H_{p}\sim\frac{1}{t}\Rightarrow a \sim t^\nu \label{nu k alpha1}
\end{equation}
\begin{equation}
c\sim t^{\frac{\alpha}{2-\alpha}}
\end{equation}
Requesting an expanding universe together with horizon criteria
lead to the following constraints on the range of possible values
of constants:
\begin{equation}
\nu>0 \ \ \ \ \ \ \ \ \  \ \nu>\frac{2}{2-\alpha}
\end{equation}
Taking now $\alpha=2$, the corresponding solutions are given by
substituting $\alpha=2$ in the relations (\ref{H}) and
(\ref{psi}). But now the time is defined as:
\begin{equation}
t\sim\ln{x^0}
\end{equation}
so that:
\begin{equation}
H_{p}\sim cons.\Rightarrow a \sim e^{\nu t}
\end{equation}
\begin{equation}
c\sim e^{\kappa t}
\end{equation}
So $\alpha=2$ leads to de Sitter expansion for cosmic scale
factor. In this case it is needed that:
\begin{equation}
\nu>0 \ \ \ \ \ \ \ \ \  \ \nu-\kappa>0
\end{equation}
Moreover another solution which leads to the power-law expansion
is:
\begin{equation}
H\sim cons. \ \ \ \ \ \ \ \ \  \psi\sim e^{\alpha x^0} \ \ \ \ \ \
F\sim\psi ^2 \label{FF}
\end{equation}
This is a special choice which is used previously by many authors
\cite{13,15,18}. The cosmic time is:
\begin{equation}
t\sim e^{-\frac{\alpha}{2}x^0}
\end{equation}
so:
\begin{equation}
H_{p}\sim\frac{1}{t}\Rightarrow a \sim t^\nu \label{ppsi}
\end{equation}
Considering both criteria pointed before, one gets the following
constraint:
\begin{equation}
\nu>0
\end{equation}
\subsubsection{$(c-\Lambda)$-dominated universe:} A mentioned before, this era corresponds
to a matter free universe for which the potential $U$ is nonzero
and has the form $U=\Lambda F$ in which $\Lambda$ is a constant.
Demanding power--law expansion for cosmic scale factor one can
easily show that the solution is:
\begin{equation}
H\sim cons. \ \ \ \ \ \ \ \ \  \psi\sim e^{\alpha x^0} \ \ \ \ \ \
F\sim\psi ^2 \label{FFf}
\end{equation}
And the cosmic time is:
\begin{equation}
t\sim e^{-\frac{1}{2}\alpha x^{0}}
\end{equation}
Here $\alpha>0$, so the speed of light is decreasing in cosmic
time.
\subsection{Palatini approach}
In the Palatini approach the metric and connections are considered as
independent fields. The variation of the action (\ref{1}) with
respect to the metric gives:
\begin{equation}
F(\Psi)(\tilde{R}_{\mu\nu}-\frac{1}{2}g_{\mu\nu}\tilde{R})=T_{\mu\nu}+ \partial_\mu\Psi\partial_\nu\Psi - {1\over 2}g_{\mu\nu}
(\partial_\alpha\Psi)^2- g_{\mu\nu} U(\Psi)
\label{ei}
\end{equation}
where $\tilde{R}_{\mu\nu}$ is Ricci tensor constructed from
connections and $\tilde{R}\equiv g^{\mu\nu}\tilde{R}_{\mu\nu}$. By
varying the action (\ref{1}) with respect to the connection, one arrives at another field equation:
\begin{equation}
\tilde{\nabla}_{\alpha}(\sqrt{-g}g^{\mu\nu}F(\Psi))=0
\end{equation}
in which $\tilde{\nabla}$ represents the covariant derivative with respect to the connection. This equation shows that the connections are Levi--Civita connections of a metric $h_{\mu\nu}$ related to $g_{\mu\nu}$ as:
\begin{equation}
h_{\mu\nu}=F(\Psi)g_{\mu\nu}
\label{h}
\end{equation}
or equivalently:
\begin{equation}
\Gamma^\lambda_{\mu\nu}={\lambda\brace\mu\nu}+\ X^\lambda_{\mu\nu}
\label{af}
\end{equation}
where:
\begin{equation}
\ X^\alpha_{\beta\gamma}=\frac{1}{2}(\delta^\alpha_\beta\partial_\gamma\ln F(\Psi)+\delta^\alpha_\gamma\partial_\beta\ln F(\Psi)-g_{\beta\gamma}g^{\alpha\delta}\partial_\delta\ln F(\Psi))
\label{aaf}
\end{equation}
is the difference between the affine connection and the Christoffel symboles (${\lambda\brace\mu\nu}$).
and finally the variation with respect to $\Psi$ gives:
\begin{equation}
\nabla_{\alpha}\nabla^{\alpha}\Psi=\frac{dU}{d\Psi}-\frac{1}{2}\frac{dF}{d\Psi}\tilde{R}
\label{Si}
\end{equation}
It is worth noting that in this case the matter energy--momentum tensor is divergence free with respect to covariant derivative defined with Levi--Civita connection of the metric, i.e. $\nabla_\mu T^\mu_\nu = 0$.  This emplies that the test particle shall move on the metric geodesic calculated using the Levi--Civita connection. An explicit proof of this point for a more general action can be found in \cite{k}.

Equation (\ref{h}) shows that the $F$--field or equivalently the $c$--field acts as a conformal factor of space--time metric,
 $g_{\mu\nu}$. Using the transformation rules of the Riemann and Ricci tensors under rescaling (\ref{h}) and then inserting
 those in  (\ref{ei}) and (\ref{Si}), we obtain:
\[
F(\Psi) (R_{\mu\nu}^{(g)}-{1\over2}g_{\mu\nu}R^{(g)}) = T_{\mu\nu}
+(1-\frac{3}{2F} {(\frac{dF}{d\Psi})}^2) (\partial_\mu\Psi\partial_\nu\Psi - {1\over 2}g_{\mu\nu}
(\partial_\alpha\Psi)^2)
\]
\begin{equation}
+\nabla_\mu\partial_\nu F(\Psi) -
g_{\mu\nu}\nabla_\alpha\nabla^\alpha F(\Psi) - g_{\mu\nu} U(\Psi)
\label{22}
\end{equation}
\begin{equation}
\nabla_{\alpha}\nabla^{\alpha}\Psi=
\frac{dU}{d\Psi}-\frac{1}{2}\frac{dF}{d\Psi}(R^{(g)}-\frac{3}{F}\nabla_{\alpha}\nabla^{\alpha}F(\Psi)+\frac{3}{2F^{2}}
(\partial_{\alpha}F)^{2})\label{ssi}
\end{equation}
where $R^{(g)}=g^{\mu\nu}R^{(g)}_{\mu\nu}$ and $R^{(g)}_{\mu\nu}$ is the Ricci tensor constructed from Christoffel symboles.
In the spatially flat FRW
space-time one can easily shown that the field equations (\ref{22}) and (\ref{ssi}) together the conservation law lead to:
\begin{equation}
3F(H^{2}+\frac{k}{a^2})=\rho+\frac{\dot{\Psi}}{2}-3H\dot{F}-\frac{3\dot{F}^{2}}{4F}+U
\label{palatini1}
\end{equation}
\begin{equation}
-2F(\dot{H}+\frac{k}{a^2})=(\rho+p)+\dot{\Psi}^{2}-H\dot{F}+\ddot{F}-\frac{3\dot{F}^{2}}{2F}
\label{palatini2}
\end{equation}
\begin{equation}
\ddot{\Psi}+3H\dot{\Psi}=3\frac{dF}{d\Psi}(\dot{H}+2H^{2}+\frac{k}{a^{2}}+\frac{3}{2}\frac{H\dot{F}}{F}+
\frac{\ddot{F}}{2F}-\frac{\dot{F}^{2}}{4F^{2}})-\frac{dU}{d\Psi}
\label{palatini3}
\end{equation}
\begin{equation}
\dot{\rho}+3H(\rho+p)=0
\end{equation}
As we did for the metric approach, one can find some exact
solutions for the spatially flat universe in two cases: $c$-dominated
and$(c-\Lambda)$-dominated. The interesting point which can be
easily shown is that demanding power-law expansion for cosmic
scale factor in the Palatini framework leads to the solutions
which are the same as the results found using the metric
formalism.

\section{Dynamics of VSL theories}
 Another way to find out some exact solutions for cosmological
models is the dynamical system method \cite{19}. In this method by
choosing some appropriate variables, one can convert the field
equations of the desired theory to a set of autonomous
differential equations. Then the critical points of the autonomous
system describe interesting exact solutions. Also this method
helps us to check the stability of the solutions. Here we consider
a class of VSL theories which are described by action (\ref{1}).
It is worth noting that this theory is the same as to a
general scalar tensor theory, but we should take into account that
the volume element is defined as $dx^{0}d^{3}x$ which is different
form the canonical volume element. So as we have explained before, since in the field equations $H$
is not the physical Hubble parameter and derivatives are with respect to $x^0$ coordinate, the dynamics of VSL theories should be different
from the scalar tensor theories. Dynamics of a general scalar tensor
theory in the Jordan frame using metric approach has been considered in \cite{20} by demanding a
background $\Lambda$CDM spatially flat cosmology.

In section 2, to find the exact solutions, we chose $F(x^{0})$ such
that the corresponding solution for cosmic scale factor was
physically interested. However, here we impose a general form for
physical Hubble parameter which is related to $\Lambda$CDM
cosmology, that is:
\begin{equation}
H_{p}(z)^{2}=H_{0}^{2}[\Omega_{0m}(1+z)^{3}+\Omega_{0r}(1+z)^{4}+\Omega_{\Lambda}]
\end{equation}
where $\Omega_{0r}=\frac{\rho_{r}}{\rho_{cr}}\simeq 10^{-4}$ ,
$\Omega_{0m}=\frac{\rho_{m}}{\rho_{cr}}\simeq 0.3$ and
$\Omega_{\Lambda}=1-\Omega_{0m}-\Omega_{0r}$, then we seek for
the corresponding solutions for scalar field. We also assume
$U(\Psi)\sim F(\Psi)^{n}$ where n is a constant \cite{21}. In
order to express equations (\ref{palatini1}),(\ref{palatini2}) and
(\ref{palatini3}) as a dynamical system of the first order
differential equations, we first write them in dimensionless form
as:
\begin{equation}
1=\Omega_{m}+\frac{\rho_{r}}{3FH^{2}}+\frac{\Psi'^{2}}{6F}+\frac{U}{3FH^{2}}-\frac{F'^{2}}{F^{2}}-\frac{F'}{F}
\label{palatini4}
\end{equation}
\begin{equation}
-2\frac{H'}{H}=\frac{\rho_{m}}{FH^{2}}+\frac{4\rho_{_{r}}}{3FH^{2}}+\frac{\Psi'^{2}}{F}-\frac{F'}{F}+\frac{H'F'}{HF}+\frac{F''}{F}
-\frac{3F'^{2}}{F^{2}} \label{palatini5}
\end{equation}
\[
\frac{\Psi''}{F}=-\frac{H'\Psi'}{HF}+\frac{1}{F}\frac{dF}{d\Psi}[\frac{3H'}{H}+6+\frac{9F'}{2F}+\frac{3H'F'}{2HF}+\frac{3F''}{2F}-\frac{3F'^{2}}{4F^{2}}
]\]
\begin{equation}
-\frac{1}{FH^{2}}\frac{dU}{d\Psi}-\frac{3\Psi'}{F}
\label{palatini6}
\end{equation}
where
$'=\frac{d}{d\ln{a}}=\frac{1}{H(x^{0})}\frac{d}{dx^{0}}=\frac{1}{H_{p}(t)}\frac{d}{dt}$.
Now we use a set of dimensionless phase-space variables
$x_{1}...x_{4}$ similar to those introduced in \cite{20}, that is:
\begin{equation}
x_{1}=-\frac{F'}{F},~~x_{2}=\frac{U}{3FH^{2}},~~x_{3}=\frac{\Psi'^{2}}{6F},~~x_{4}=\frac{\rho_{r}}{3FH^{2}}
\end{equation}
Now defining $\Omega_{m}=\frac{\rho_{m}}{3FH^{2}}$ we write
equations (\ref{palatini4}) and (\ref{palatini5}) as:
\begin{equation}
\Omega_{m}=1-x_{4}-x_{3}-x_{2}-x_{1}+\frac{1}{4}x_{1}^{2}
\end{equation}
\begin{equation}
x_{1}'=3+2\frac{H'}{H}+x_{4}+3x_{3}-3x_{2}-(2+\frac{H'}{H})x_{1}+\frac{1}{4}x_{1}^{2}
\label{palatini7}
\end{equation}
Differentiating $x_{4}$ and $x_{2}$ with respect to $\ln{a}$ gives:
\begin{equation}
x_{4}'=x_{4}[x_{1}-4-2\frac{H'}{H}] \label{palatini8}
\end{equation}
\begin{equation}
x_{2}'=x_{2}[x_{1}(1-n)-2\frac{H'}{H}] \label{palatini9}
\end{equation}
where $n=\frac{F}{U}\frac{dU}{dF}$. Finally differentiating
$x_{3}$ with respect to $\ln{a}$ and using (\ref{palatini6}), we have:
\begin{equation}
x_{3}'=-2(\frac{H'}{H}+3)x_{3}+\frac{1}{2}x_{1}(x_{1}+x_{4}+5x_{3}+(2n-3)x_{2}-1)-\frac{1}{8}x_{1}^{3}
\label{palatini10}
\end{equation}
On the other hand one can easily verify that
$\frac{H'}{H}=\frac{H_{p}'}{H_{p}}+\frac{1}{4}x_{1}$, so by
substituting this relation to equations (\ref{palatini7}),
(\ref{palatini8}), (\ref{palatini9}) and (\ref{palatini10}) we
have:
\begin{equation}
x_{1}'=3-(\frac{3}{2}+\frac{H_{p}'}{H_{p}})x_{1}-3x_{2}+3x_{3}+2\frac{H_{p}'}{H_{p}}+x_{4}
\label{palatini11}
\end{equation}
\begin{equation}
x_{2}'=x_{2}[x_{1}(\frac{1}{2}-n)-2\frac{H_{p}'}{H_{p}}]
\label{palatini12}
\end{equation}
\begin{equation}
x_{3}'=-2(\frac{H_{p}'}{H_{p}}+3)x_{3}+\frac{1}{2}x_{1}(x_{1}+x_{4}+4x_{3}+(2n-3)x_{2}-1)-\frac{1}{8}x_{1}^{3}
\label{palatini13}
\end{equation}
\begin{equation}
x_{4}'=x_{4}[\frac{1}{2}x_{1}-4-2\frac{H_{p}'}{H_{p}}]
\label{palatini14}
\end{equation}
These equations describe the cosmological dynamics of the VSL
theory in Palatini formalism. By a completely similar procedure
one can write the field equations of the metric formalism as
follows:
\begin{equation}
x_{1}'=-3-x_{1}(\frac{3}{2}+\frac{H_{p}'}{H_{p}})-3x_{2}+x_{4}+2x_{3}+\frac{3}{4}x_{1}^{2}+2\frac{H_{p}'}{H_{p}}
\label{metric1}
\end{equation}
\begin{equation}
x_{2}'=x_{2}[x_{1}(\frac{1}{2}-n)-2\frac{H_{p}'}{H_{p}}]
\label{metric2}
\end{equation}
\begin{equation}
x_{3}'=x_{1}[\frac{1}{2}x_{3}-2-\frac{H_{p}'}{H_{p}}+nx_{2}]-6x_{3}-2\frac{H_{p}'}{H_{p}}x_{3}-\frac{1}{4}x_{1}^{2}
\label{metric3}
\end{equation}
\begin{equation}
x_{4}'=x_{4}[\frac{1}{2}x_{1}-4-2\frac{H_{p}'}{H_{p}}]
\label{metric4}
\end{equation}

It is important to note that $\frac{H_{p}'}{H_{p}}$ is not always
constant so the dynamical equations of the metric and Palatini formalisms
are not autonomous and we can not find the critical points in any
regime. On the other hand we know that $\frac{H_{p}'}{H_{p}}$ for
$\Lambda$CDM background is approximately constant in the matter,
radiation and de Sitter eras and so we can use the dynamical system
method for these eras \cite{20}. The critical points and their
corresponding eigenvalues are shown in  Table(\ref{tab:m}) for the metric formalism
and in Table(\ref{tab:p}) for the Palatini formalism.
\vspace{0pt}
\begin{table*}[h]
\begin{center}

\caption{The critical points of the system (\ref{metric1}),
(\ref{metric2}), (\ref{metric3}) and (\ref{metric4})}
\begin{tabular}{ccccccccc}
\hline \hline\\
\vspace{1pt} \textbf{Era}&  \textbf{Cp $(x_{1},x_{2},x_{3},x_{4})$}   \hspace{7pt}& \hspace{7pt} \textbf{Eigenvalues} \hspace{7pt} \\
\hline
\vspace{1pt}                      &$R_1$(0,~0,~0,~1)&(1,~-1/2,~-2,~4)\\
\vspace{1pt}                      &$R_2$(2,~0,~-1,~0)&(1,~2,~1/2,~5-2$n$)\\
\vspace{1pt}Radiation   &$R_3$(4/3,~0,~-1/3,~0)&(5/3,~-1/2,~2/3,~$\frac{14}{3}-\frac{4n}{3}$)\\
\vspace{1pt} $\frac{H_{p}'}{H_{p}}=-2$&$R_4$(-2,~0,~-1/3,~0)&(-1,$\frac{\sqrt{41}-11}{4}$,$-\frac{\sqrt{41}+11}{4}$,3+2$n$ )\\
\vspace{1pt}                      &$R_5$($\frac{8}{2n-1}$,$\frac{(4n^{2}-24n+35)}{3(2n-1)^{2}}$,$\frac{4(2n^{2}-9n-2)}{3(2n-1)^{2}}$,0)&($\frac{4}{2n-1}$,$\frac{2n+3}{2n-1}$,$\frac{10n-29-A}{4-8n}$,$\frac{10n-29+A}{4-8n}$)\\
\hline
\vspace{1pt}                      &$M_1$(0,~0,~0,~0)&(3,~-1,~$\frac{\sqrt{3}-3}{2}$,~$-\frac{\sqrt{3}+3}{2}$)\\
\vspace{1pt}  Matter                    &$M_2$(2,~0,~-1,~0)&(0,~0,~1,$4-2n$)\\
\vspace{1pt} $\frac{H_{p}'}{H_{p}}=-\frac{3}{2}$  &$M_3$($\frac{6}{2n-1}$,$\frac{2(4+n^{2}-4n)}{(2n-1)^{2}}$,$\frac{2n^{2}-8n-1}{(2n-1)^{2}}$,0)&(-1,$\frac{3}{2n-1}$,$\frac{(2-n)(-3+C)}{2n-1}$,$\frac{(2-n)(3+C)}{2n-1}$)\\
\hline
\vspace{1pt}                      &$\Lambda_1$(0,~1,~0,~0)&(-4,-3,$\frac{-9+\sqrt{33+96n}}{4}$,$\frac{-9-\sqrt{33+96n}}{4}$)\\
\vspace{1pt}                      &$\Lambda_2$(6,~0,~-7,~0)&(-1,$\frac{9+\sqrt{33}}{4}$,$\frac{9-\sqrt{33}}{4}$,$3-6n$)\\
\vspace{1pt}De Sitter   &$\Lambda_3$(2,~0,~-1,~0)&(-3/2,~-3,~-2,~$1-2n$)\\
\vspace{1pt} $\frac{H_{p}'}{H_{p}}=0$&$\Lambda_4$(4,~0,~-3,~0)&(3/2,~-2,~-1,~$2-4n$ )\\
\hline \hline
\label{tab:m}
\end{tabular}
\end{center}
\end{table*}

Considering Table(\ref{tab:p}), we see that the only physically interested
coordinates are $R_{1},R_{5},M_{1},M_{3}$ and $\Lambda_{1}$, for
other points $x_{3}<0$ for any $n$. For $n<\frac{1}{2}$,
$\Lambda_{1},M_{3}$ and $R_{5}$ are attractors, but demanding
positive values for $x_{3}$ imposes another range to n, say
$-0.93<n<-0.13$. Note that $A=\sqrt{281+128n^{3}+924n-732n^{2}}$
and $C=\sqrt{3+12n}$. It should be noted that $R_{1},M_{1}$ and
$\Lambda_{1}$ correspond to general relativity with constant speed of light.
Also it is worth noting that these results are different from those
obtained in \cite{20} for similar scalar tensor theory.
Considering the critical points of Table3 we see that $M_{3}'$ is
an attractor if $-\frac{1}{4}<n<\frac{1}{2}$. Also in the de Sitter
era $\Lambda_{1}'$ for $n<\frac{1}{2}$ and $\Lambda_{3}'$ for
$n>\frac{1}{2}$ are attractors. The noteworthy feature of the Palatini
VSL theory is that there exists no attractor critical point in
the radiation dominated universe and unlike the metric formalism, for
all critical points, $x_{3}$ can be a positive quantity. However,
in the Palatini formalism we have three critical points $R_{3}'$,
$M_{1}'$ and $\Lambda_{3}'$, which are meaningless because for
them $F'(\Psi)\neq0$ while $\Psi'=0$. It is also important to
mention that again $R_{2}',M_{2}'$ and $\Lambda_{1}'$ correspond
to general relativity with constant speed of light.

\vspace{0pt}
\begin{table*}[h]
\begin{center}
\caption{The critical points of the system (\ref{palatini11}),
(\ref{palatini12}), (\ref{palatini13}) and (\ref{palatini14})}
\begin{tabular}{ccccccccc}
\hline \hline\\
\vspace{1pt} \textbf{Era}&  \textbf{Cp $(x_{1},x_{2},x_{3},x_{4})$}   \hspace{7pt}& \hspace{7pt} \textbf{Eigenvalues} \hspace{7pt} \\
\hline
\vspace{1pt}                      &$R_1'$($\frac{8}{2n-1}$,$\frac{4n^{2}-24n+35}{3(2n-1)^{2}}$,$\frac{4(2n^{2}-9n+10)}{3(2n-1)^{2}}$,0)&(0,$\frac{2n+3}{2n-1}$,$\frac{10n-29-A}{4-8n}$,$\frac{10n-29+A}{4-8n}$)\\
\vspace{1pt}                      &$R_2'$(0,~0,~0,~1)&(-2,~4,~$\frac{1+\sqrt{17}}{4}$,~$\frac{1-\sqrt{17}}{4}$)\\
\vspace{1pt}Radiation             &$R_3'$(2,~0,0,~0)&(0,~1/2,~2,~$5-2n$)\\
\vspace{1pt} $\frac{H_{p}'}{H_{p}}=-2$&$R_4'$(4/3,~0,~1/9,~0)&(0,~-1/2,~5/3,~$\frac{14}{3}(1-n)$)\\
\vspace{1pt}                      &$R_5'$(-2,~0,~2/3,~0)&(0,$\frac{\sqrt{41}-11}{4}$,~$-\frac{\sqrt{41}+11}{4}$,~$3+2n$)\\
\hline
\vspace{1pt}                      &$M_1'$(2,~0,~0,~0)&(0,~1,~-1,~$4-2n$)\\
\vspace{1pt}  Matter                    &$M_2'$(0,~0,~0,~0)&(3,~-1,~$\frac{\sqrt{3}-3}{2}$,$-\frac{\sqrt{3}+3}{2}$)\\
\vspace{1pt} $\frac{H_{p}'}{H_{p}}=-\frac{3}{2}$  &$M_3'$($\frac{6}{2n-1}$,$\frac{2(4+n^{2}-4n)}{(2n-1)^{2}}$,$\frac{2(4+n^{2}-4n)}{(2n-1)^{2}}$,0)&(-1,$\frac{3}{2n-1}$,$\frac{(2-n)(-3+C)}{2n-1}$,$\frac{(2-n)(3+C)}{2n-1}$)\\
\hline
\vspace{1pt}                      &$\Lambda_1'$(0,~1,~0,~0)&(-4,-3,$\frac{-9+\sqrt{33+96n}}{4}$,$\frac{-9-\sqrt{33+96n}}{4}$)\\
\vspace{1pt}                      &$\Lambda_2'$(6,~0,~2,~0)&(-4,$\frac{9+\sqrt{33}}{4}$,$\frac{9-\sqrt{33}}{4}$,$3-6n$)\\
\vspace{1pt}De Sitter   &$\Lambda_3'$(2,~0,~0,~0)&(-3/2,~-4,~-2,~$1-2n$)\\
\vspace{1pt} $\frac{H_{p}'}{H_{p}}=0$&$\Lambda_4'$(4,~0,~1,~0)&(3/2,~-4,~-1,~$2-4n$ )\\
\hline \hline
\label{tab:p}
\end{tabular}
\end{center}
\end{table*}

It is interesting to note that for any solutions which are
extracted from the analysis of the critical points, the non-minimal
coupling coefficient takes the form $F(\Psi)\sim
(\Psi-\Psi_{0})^{2}$. For any regime and critical point, the speed
of light takes the form $c(t)\sim a(t)^{-x_{1}/4}$ in both
formalisms. The scalar field takes the form $\Psi(t)=\lambda
t^{-x_{1}/4}+\Psi_{0}$ and the horizon criteria is $x_{1}>4$ in
the radiation dominated era. In the matter dominated era, the scalar field is
$\Psi(t)=\lambda t^{-x_{1}/3}+\Psi_{0}$ and the horizon criteria
takes the form $x_{1}>2$.  On the other hand considering the above
tables, we see that the critical points in the matter and radiation eras do
not satisfy the horizon criteria. However, in the de Sitter era the
horizon criteria is satisfied by any physical critical point and
the solution for $\Psi$ is $\Psi(t)=\lambda
e^{-x_{1}t/4}+\Psi_{0}$.

\section{Conclusions}
In this letter, we have considered a class of VSL theories which
are described by the action (\ref{1}), in both metric and Palatini
formalisms. In section 2 we have shown that this model exhibits a
power-law or de Sitter expansion for the cosmic scale factor. In
any case applying both formalisms, we have shown that $F(\Psi)$ is
a quadratic function in terms of the dynamical scalar field
$\Psi$. This is a particular case emerged by requiring the
existence of the Noether symmetry for any cosmological point-like
lagrangian of a general scalar tensor theory \cite{16}. On the other hand, one aspect of our model is if the scalar field can be interpreted as dark energy as well as variable speed of light. Many authors studied for example the change of fine structure constant based on quintessence \cite{olive}. To give a look at this point here, note that in the metric
formalism, using equations (\ref{H2}) and (\ref{dotH}), we define
$\omega_{eff}$ as follows:
\begin{equation}
\omega_{eff}=\frac{P_{eff}}{\rho_{eff}}=\frac{\dot{\Psi}^{2}+2\ddot{F}+4H\dot{F}-2U}{\dot{\Psi}^{2}-6H\dot{F}+2U}
\label{omegametric}
\end{equation}
The choice of $\omega_{eff}=-1$ leads to the following equation:
\begin{equation}
\dot{\Psi}^{2}-H\dot{F}+\ddot{F}=0 \label{omegametric1}
\end{equation}
In the matter--radiation free universe, by using equations (\ref{H2}),(\ref{dotH}) and (\ref{omegametric1}) we have:
\begin{equation}
6FH_{0}^{2}+5H_{0}\dot{F}+\ddot{F}-2U=0
 \label{omegametric2}
\end{equation}
where $H=H_{0}$ is a constant. By using these equations one can
verify that the coupling coefficient takes the form $F(\Psi)\sim
\Psi^{2}$ in both c--dominated and $\Lambda$--dominated
universe. As an example of this point, consider a $\Lambda$--dominated universe. The
equation (\ref{omegametric2}) leads to the following solution for
the coupling function:
\begin{equation}
F \sim e^{\frac{-1}{2}(5H_{0}+\sqrt{H^{2}_{0}+8\Lambda})x^{0}}
\label{ccc}
\end{equation}
And by substituting this result in (\ref{omegametric1}) we obtain:
\begin{equation}
\Psi \sim
e^{\frac{-1}{4}(5H_{0}+\sqrt{H^{2}_{0}+8\Lambda})x^{0}}
\end{equation}
 So  $F(\Psi)\sim
\Psi^{2}$. Similarly, in the Palatini formalism, we define
$\omega_{eff}$ as follows:
\begin{equation}
\omega_{eff}=\frac{P_{eff}}{\rho_{eff}}=\frac{\dot{\Psi}^{2}+2\ddot{F}+4H\dot{F}-\frac{3\dot{F}^{2}}{2F}-2U}{\dot{\Psi}^{2}-6H\dot{F}-\frac{3\dot{F}^{2}}{2F}+2U}
\end{equation}
Again, $\omega_{eff}=-1$ leads to the following equation:
\begin{equation}
\dot{\Psi}^{2}-H\dot{F}+\ddot{F}-\frac{3\dot{F}^{2}}{2F}=0
 \label{omegappalatini1}
\end{equation}
By considering the modified Friedman equations of this formalism, equations (\ref{palatini1}) and (\ref{palatini2}),
in the matter--radiation free universe, one can see that the coupling function satisfying the condition (\ref{omegametric2}). So by using the equations
(\ref{omegametric2}) and (\ref{omegappalatini1}), we see that
$F(\Psi)\sim \Psi^{2}$ in both c--dominated and
$\Lambda$--dominated universe. So the consistency with Nother symmetry and having constant $\omega_{eff}$,
specially $\omega_{eff}=-1$, are both satisfied for $F(\Psi)\sim \Psi^{2}$. But it should be stressed here that
unlike the usuall $\Lambda$CDM or quintessence models, the expression  $\omega_{eff}=-1$
dose not mean necessarily an accelerated universe (i.e. $\frac{d^2a}{dt^2}>0$).
To clarify this point let us combine (\ref{H2}) and (\ref{dotH}) in the metric formalism
(or (\ref{palatini1}) and (\ref{palatini2}) in the Palatini formalism) as follows:
\begin{equation}
\frac{\ddot{a}}{a}=-\frac{1}{6}[(1+3\omega)\rho+(1+3\omega_{eff})\rho_{eff}]
\end{equation}
Converting the derivatives in terms of $x^{0}$ to derivatives
with respect to the cosmic time, for a matter--radiation free universe, we have:
\begin{equation}
\frac{1}{a}\frac{d^2a}{dt^2}=\frac{1}{4}\frac{1}{F}\frac{dF}{dt}H_{p}-\frac{(1+3
\omega_{eff})}{6F^{1/2}}\rho_{eff} \label{adotdot}
\end{equation}
Note that this equation is correct in both formalisms with only different expression of $\rho_{eff}$ so the following results are common for them. From this equation we see that the evolution of the speed of light has
a crucial role in constructing an inflationary universe. If the speed
of light is decreasing in cosmic time, then the necessary condition
for $\frac{d^2a}{dt^2}>0$ is $\omega_{eff}<-\frac{1}{3}$, but this is not
sufficient condition. On the other hand, if the speed of light is
increasing in time, then the sufficient condition for $\frac{d^2a}{dt^2}>0$ is
$\omega_{eff}<-\frac{1}{3}$, but this is not necessary condition. In
the other word, by a constant $\omega_{eff}$, the sign of $\frac{d^2a}{dt^2}$
can change with time. For a simple example, consider the solution
(\ref{ccc}), where $H(x^{0})=H_{0}$ and $\omega_{eff}=-1$. By
using equation (\ref{adotdot}) we have:

\begin{equation}
\frac{1}{a}\frac{d^2a}{dt^2}= H_{p}^{2}(1- \alpha t^{2}).
\end{equation}
where $\alpha$ is a positive constant.

We recall that the action (\ref{1}) which is used in this paper, allows for a dynamical gravitational constant as well as dynamical speed of light. The only difference is that we should take $x^0$ coordinate as a time--like coordinate in the latter case. So the equations of variable speed of light and variable gravitational constant scalar--tensor cosmology are not similar when we write them in terms of the physical time and physical Hubble parameter. To clarify better, starting action (\ref{1}) as a variable gravitational constant theory and then applying it to cosmology, considering a matter--radiation free universe, it can be easily shown that the first term in equation (\ref{adotdot}) is absent in this case. In this theory $\omega_{eff}<- \frac{1}{3}$ leads to an inflationary universe if the matter--radiation density is negligible.

In section 3 we have considered the dynamics of the VSL theories
in order to find out other exact solutions. We have constructed
the cosmological dynamical system which is constrained to obey the
$\Lambda$CDM cosmic history. Also by considering the corresponding
critical points, we have shown that for both formalisms, variable
speed of light takes the form $c\sim a^{-\frac{x_{1}}{4}}$ in each
era. But the time-dependence of dynamical scalar field is
different in the radiation and matter eras and it has the forms $
\Psi \sim t^{-\frac{x_{1}}{4}}$ and $ \Psi \sim
t^{-\frac{x_{1}}{3}}$, respectively.

By calculating the variables $x_{1},...,x_{4}$  for exact
solutions of the c-- dominated universe which have been obtained
in section 2 and comparing them with the results of the section 3,
one can show that they are not critical points in metric
formalism. But by choosing appropriate values for $\alpha,
\upsilon$ and $k$ in the relations (\ref{nu k
alpha1})--(\ref{ppsi}), they can correspond to the critical points
$R'_{4}$, $R'_{5}$, $\Lambda'_{2}$ and $\Lambda'_{4}$ in Palatini
formalism.

It is worth to note that for any solution which are extracted
from the analysis of critical points, the non-minimal coupling
coefficient takes the form $F(\Psi)\sim (\Psi-\Psi_{0})^{2}$. Also
unlike the metric approach, there is no attractor critical point
in the radiation dominated era for the Palatini approach. It is also
important to note that critical points in the matter and radiation eras
do not satisfy the horizon criteria. However in the de Sitter era the
horizon criteria is satisfied by any physical critical point and
the scalar field has the exponential dependence to cosmic time.

\[ \]
{\bf Acknowledgements:}

This work is partly supported by a grant from university of Tehran
and partly by a grant from center of excellence of department of
physics on the structure of matter.

\end{document}